\documentclass[aps,prd,showpacs,twocolumn,amsmath]{revtex4}
\usepackage{graphicx}
\usepackage{epsfig}
\usepackage{epsf}
\usepackage{dcolumn}
\usepackage{bm}
\usepackage{multirow}
\def \ldc {\Lambda^+_c}
\def \half {{1\over 2}}
\def \ee {e^+e^-}

\def \jp {J/\psi}
\newcommand{\ket}[1]{|#1\rangle}
\newcommand{\cg}[2]{\langle #1|#2\rangle}

\newcommand{\djmn}[3]{D^{#1}_{#2}(#3)}
\newcommand{\myvec}[1]{{\bf #1}}
\begin{document}

\title{Coherent helicity amplitude for sequential decays}

\author{Hong Chen$^a$, Rong-Gang Ping$^{b,c}$\\
a) School of Physical Science and Technology, Southwest University, Chongqing, 400715, China\\
b) Institute of High Energy Physics, Chinese Academy of Sciences,\\
 P.O. Box 918(1), Beijing 100049, China\\
c) University of Chinese Academy of Science, Beijing 100049, China
\date{ }}

\email[]{pingrg@ihep.ac.cn}

\begin{abstract}
We present a derivation of coherent helicity amplitudes for a particle decaying into multifinal states with nonzero spins. The results show that the coherent amplitudes introduce additional rotations to transform the helicities into a consistent helicity system, which allows us to add helicity amplitudes for different decay chains coherently. These rotations may have significant effects on the interference between the decay chains in the partial wave analysis.
\end{abstract}

\pacs{13.25.-k, 13.60.Le, 13.75.Lb, 14.40.-n}
\maketitle
\section{Introduction}
The amplitude for a particle decaying into final states with nonzero spins can be formalized in different descriptions, such as the covariant description \cite{rarita}, the projection operators of arbitrary spin \cite{fronsdal}, and the helicity formalism originally developed by Jacob and Wick \cite{jacobi}. The structure of helicity formalism can be decomposed into the angular distribution and kinematic dependence part. This feature facilitates the determination of the parent particle properties, e.g., spin and parity quantum numbers in experiment, by analyzing the angular distribution of daughter particle. The relationships of the helicity formalism to the covariant and operator formalism were developed in Refs. \cite{chung1,chung2,chung3}.

The helicity amplitude for a sequential decay can be formalized by multiplying the amplitude of each decay chain together in a straightforward way \cite{chung3,pingrg1}. Then the partial decay rate is calculated by taking the sum of the helicity amplitude squared over the helicities of final states and taking the average over the parent particle spin. However, if there are multidecay chains into the same final states, their helicities may be defined in the different helicity system. This makes the helicities inconsistent when taking the sum of the decay chain amplitudes to calculate the partial decay rate. Such an issue of coherent helicity amplitude has been addressed in recent analyses \cite{lhcb,belle}.

The purpose of this paper is to present a derivation of coherent helicity amplitude for sequential decays. The idea is to relate the matrix element of the decay amplitude defined in the helicity basis to that defined in the canonical basis. This allows one to add the amplitude for different decay chains coherently. The coherent helicity amplitude should give the same decay rate as that calculated with the canonical basis. This requirement automatically introduces an additional rotation to transform the helicities into the same reference.

We start the discussion with two-body decays in Sec. \ref{sectionI}. Although the results in this section are well known, it is convenient to review them for use in later sections and to establish our notations. In Sec. \ref{sectionII}, we demonstrate how to construct the coherent helicity amplitude for two different decay chains in three-body decays. Two examples in three-body decays are shown in Sec. \ref{sectionIII}. For a specific decay, the helicity amplitude can be further reduced with the conservation requirements, such as the parity conservation for strong and electromagnetic decays, or the identical particle symmetry, and can be related to the covariant helicity-coupling amplitudes as discussed in Refs. \cite{chung1,chung2,chung3}.

\section{Two-body decay \label{sectionI}}
Considering a particle with spin $J$ decaying into two-body final states with spin and helicity defined in Table \ref{table1}, states of angular momentum $J$ may be constructed in the center-of-mass (CM) system of the parent particle as \cite{jacobi,chung1,chung2,chung3}
\begin{equation}
|JM\lambda_1\lambda_2\rangle = a \int d\Omega \djmn{J*}{M,\lambda_1-\lambda_2}{\phi,\theta,0}\ket{\phi\theta\lambda_1\lambda_2},
\end{equation}
where $a$ is a normalization constant, $\Omega(\theta,\phi) $ is the solid angle of final states, $\ket{\phi\theta\lambda_1\lambda_2}$ is the two-particle state in the helicity basis, defined as
\begin{eqnarray}
\ket{\theta\phi\lambda_1\lambda_2}&=&a U[R(\Omega)]\{U[L_z(p)]\ket{s_1\lambda_1}\nonumber\\
&\times&U[L_{-z}(p)]\ket{s_2-\lambda_2}\},
\end{eqnarray}
where $R(\Omega)$ denotes a rotation which carries the $z$ axis into the direction of momentum ${\bf p}$, and $L_{\pm z}(p)$ is the boost along the  $\pm z$ axis, $U[...]$ denotes operator.
The helicity states are related to the canonical states by
\begin{eqnarray}
\ket{\theta\phi\lambda_1\lambda_2}&=&aU[L(\myvec{p})]U[R(\Omega)]\ket{s_1\lambda_1}\nonumber\\
&\times&U[L(-\myvec{p})]U[R(\Omega)]\ket{s_2-\lambda_2}\nonumber\\
&=&\sum_{m_1m_2}\djmn{s_1}{m_1,\lambda_1}{\phi,\theta,0}\nonumber\\
&\times&\djmn{s_2}{m_2,-\lambda_2}{\phi,\theta,0}\ket{\theta\phi m_1m_2}.
\end{eqnarray}

\begin{table}
\caption{Variables defined for the two-body decay $J\to s_1+s_2$.\label{table1}}
\begin{tabular}{lccc}
\hline\hline
      & Parent & Daughter 1 & Daughter 2\\\hline
Spin  &  $J$   & $s_1$      & $s_2$\\
Spin $z$ projection & $M$ & $m_1$ & $m_2$\\
Helicity &---             & $\lambda_1$ & $\lambda_2$\\
Momentum & ${\bf 0}$ & ${\bf p}$  & ${\bf -p}$\\
\hline\hline
\end{tabular}
\end{table}

Using above equations, the element to project helicity states onto canonical states reads
\begin{eqnarray}
\cg{\theta\phi m_1m_2}{JM\lambda_1\lambda_2}&=&a \djmn{J*}{M,\lambda_1-\lambda_2}{\phi,\theta,0}\nonumber\\
&\times&\djmn{s_1}{m_1,\lambda_1}{\phi,\theta,0}\djmn{s_2}{m_2,-\lambda_2}{\phi,\theta,0}.\nonumber\\
\end{eqnarray}
The matrix element of amplitude for the two-body decay defined in canonical basis is related to that defined in the helicity basis by
\begin{eqnarray}
&&\mathcal{A}^J_{c}(\theta,\phi; m_1,m_2)=\langle \theta\phi m_1m_2| \mathcal{M}|JM\rangle\nonumber\\
&=&\sum_{\lambda_1,\lambda_2}\cg{\theta\phi m_1m_2}{JM\lambda_1\lambda_2}\cg{JM\lambda_1\lambda_2}{\mathcal{M}|JM}\nonumber\\
&=&a\sum_{\lambda_1\lambda_2} \djmn{J*}{M,\lambda_1-\lambda_2}{\phi,\theta,0}
\djmn{s_1}{m_1,\lambda_1}{\phi,\theta,0}\nonumber\\
&\times&\djmn{s_2}{m_2,-\lambda_2}{\phi,\theta,0}F^{J}_{\lambda_1,\lambda_2},
\end{eqnarray}
where $F^{J}_{\lambda_1,\lambda_2}=\cg{JM\lambda_1\lambda_2}{\mathcal{M}|JM}$ is helicity amplitude. Due to the orthogonality of Wigner-$D$ function, the decay rate $d\Gamma$ is proportional to
\begin{eqnarray}
d\Gamma &\propto& \sum_{\bar Mm_1m_2}|\mathcal{A}^J_{c}(\theta,\phi; m_1,m_2)|^2d\Omega\nonumber\\
&=&a^2\sum_{\bar M\lambda_1\lambda_2} \djmn{J*}{M,\lambda_1-\lambda_2}{\phi,\theta,0} \djmn{J}{M,\lambda_1-\lambda_2}{\phi,\theta,0}\nonumber\\
&\times&|F^J_{\lambda_1,\lambda_2}|^2d\Omega.
\end{eqnarray}
This equation indicates that the decay rate is calculated in the helicity basis by taking the helicity amplitude as
\begin{equation}
\mathcal{A}^J_h(\theta,\phi,\lambda_1,\lambda_2)=a \djmn{J*}{M,\lambda_1-\lambda_2}{\phi,\theta,0} F^J_{\lambda_1,\lambda_2}.
\end{equation}

\section{Sequential decay  \label{sectionII}}
\begin{figure}[htbp]
\includegraphics[height=5cm]{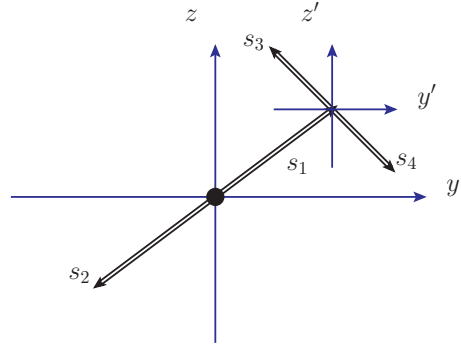}%
\caption{The orientation of the coordinate systems associated with the sequential decay $J\to s_1+s_2$, $s_1\to s_3+s_4$. \label{fig1}}
\end{figure}

We take a three-body decay, $J\to s_2+s_3+s_4$, as an example to illustrate how to calculate the helicity amplitude for sequential decays. The solid angles for each decay are defined in the rest frame of the parent particle. Considering two different decay chains I: $J\to s_1+s_2$, $s_1\to s_3+s_4$ as shown in Fig. \ref{fig1},  and II: $J\to s_0+s_3$, $s_0\to s_2+s_4$, the solid angles and kinematic variables are defined in Tables \ref{chaina} and \ref{chainb}.

\begin{table*}
\caption{Variables defined for the decay chain I, $J\to s_1+s_2$, with $s_1\to s_3+s_4$. \label{chaina}}
\begin{tabular}{lccc|lccc}
\hline\hline
 &\multicolumn{3}{c}{First decay} & \multicolumn{3}{|c}{Second decay} \\\hline
      & Parent & Daughter 1 & Daughter 2& Parent & Daughter 1 & Daughter 2\\\hline
Spin  &  $J$   & $s_1$      & $s_2$&$s_1$&$s_3$&$s_4$\\
Spin $z$ projection & $M$ & $m_1$ & $m_2$&$m_1$&$m_3$&$m_4$\\
Helicity &---             & $\lambda_1$ & $\lambda_2$&$\lambda_1,\lambda'_1$&$\lambda_3$&$\lambda_4$\\
Helicity amplitude & &$F^J_{\lambda_1,\lambda_2}$ &   & &$F^{s_1}_{\lambda_3,\lambda_4}$&&\\
Solid angle &  & $\Omega_1(\theta_1,\phi_1)$ & & & $\Omega_2(\theta_2,\phi_2)$& \\
\hline\hline
\end{tabular}
\end{table*}

\subsection{Amplitude for single decay chain}
Let us consider the first decay chain I, where the amplitude in the canonical state, $\mathcal{A}_\textrm{I}^\textrm{a}$, can be expressed with the helicity amplitude as

\begin{eqnarray}
&&\mathcal{A}_\textrm{I}^\textrm{a}(\theta_1,\phi_1;M,m_1,m_2)=\sum_{\lambda_1,\lambda_2}
\djmn{s_1}{m_1,\lambda_1}{\phi_1,\theta_1,0}\nonumber\\
&\times&\djmn{s_2}{m_2,-\lambda_2}{\phi_1,\theta_1,0}
\djmn{J*}{M,\lambda_1-\lambda_2}{\phi_1,\theta_1,0}F^J_{\lambda_1,\lambda_2}.
\end{eqnarray}

The amplitude for the second decay is

\begin{eqnarray}
&&\mathcal{A}_\textrm{I}^\textrm{b}(\theta_2,\phi_2;m_1,m_3,m_4)=\sum_{\lambda_3,\lambda_4}
\djmn{s_3}{m_3,\lambda_3}{\phi_2,\theta_2,0}\nonumber\\
&\times&\djmn{s_4}{m_4,-\lambda_4}{\phi_2,\theta_2,0}
\djmn{s_1*}{m_1,\lambda_3-\lambda_4}{\phi_2,\theta_2,0}F^{s_1}_{\lambda_2,\lambda_2}.
\end{eqnarray}

The rotation $\djmn{s_1*}{m_1,\lambda_3-\lambda_4}{\phi_2,\theta_2,0}$ carries the $z$ axis into the direction of $s_3$ momentum, and we decompose it into two successive rotations. We first rotate the $z$ axis into the direction of $s_1$ in its parent CM system, and then rotate into the direction of $s_3$ momentum at the $s_1$ CM system by the helicity angle $\bar\Omega_2(\bar\theta_2,\bar\phi_2)$. It follows the multiplication rule
\begin{eqnarray}
\djmn{s_1*}{m_1,\lambda_3-\lambda_4}{\phi_2,\theta_2,0}&=&\sum_\delta \djmn{s_1*}{m_1,\delta}{\phi_1,\theta_1,0}\nonumber\\
&\times&\djmn{s_1*}{\delta,\lambda_3-\lambda_4}{\bar\phi_2,\bar\theta_2,0}.
\end{eqnarray}

The amplitude in the canonical state for the sequential decay is

\begin{eqnarray}
&&\mathcal{A}_\textrm{I}(\theta_1,\theta_2,\phi_1,\phi_2; J,M,m_2,m_3,m_4)\nonumber\\
&=&\sum_{m_1}\mathcal{A}_\textrm{I}^\textrm{a}(\theta_1,\phi_1;M,m_1,m_2)\mathcal{A}_\textrm{I}^\textrm{b}(\theta_2,\phi_2;m_1,m_3,m_4)\nonumber\\
&=&\sum_{\lambda_1,\lambda_2,\lambda_3,\lambda_4}\djmn{s_2}{m_2,-\lambda_2}{\phi_1,\theta_1,0}\djmn{s_3}{m_3,\lambda_3}{\phi_2,\theta_2,0}\nonumber\\
&\times&\djmn{s_4}{m_4,-\lambda_4}{\phi_2,\theta_2,0}\djmn{J*}{M,\lambda_1-\lambda_2}{\phi_1,\theta_1,0}\nonumber\\
&\times&\djmn{s_1*}{\lambda_1,\lambda_3-\lambda_4}{\bar\phi_2,\bar\theta_2,0}
F^J_{\lambda_1,\lambda_2}F^{s_1}_{\lambda_3,\lambda_4}.
\end{eqnarray}

The orthogonality relation, $$\sum_{m_1}\djmn{s_1}{m_1,\lambda_1}{\phi,\theta_1,0}\djmn{s_1*}{m_1,\delta}{\phi,\theta_1,0}=\delta_{\lambda_1,\delta},$$ is used in the above equation.
The decay rate for the sequential decay is proportional to

\begin{eqnarray}
d\Gamma &\propto& \sum_{\mbox{\tiny$\begin{array}{c}\bar M,m_2,m_3\\m_4,\lambda_1,\lambda'_1\\
\lambda_2,\lambda_3,\lambda_4\end{array}$}}|\mathcal{A}_\textrm{I}(\theta_1,\theta_2,\phi_1,\phi_2; J,M,m_2,m_3,m_4)|^2\nonumber\\
&=&\sum_{\mbox{\tiny$\begin{array}{c}\bar M,\lambda_1,\lambda'_1\\~\lambda_2,\lambda_3,\lambda_4\end{array}$}}\djmn{J*}{M,\lambda_1-\lambda_2}{\phi_1,\theta_1,0}
\djmn{J}{M,\lambda'_1-\lambda_2}{\phi_1,\theta_1,0}\nonumber\\
&\times&\djmn{s_1*}{\lambda_1,\lambda_3-\lambda_4}{\bar\phi_2,\bar\theta_2,0}\djmn{s_1}{\lambda'_1,\lambda_3-\lambda_4}{\bar\phi_2,\bar\theta_2,0}\nonumber\\
&\times&F^{J*}_{\lambda'_1,\lambda_2}F^{J}_{\lambda_1,\lambda_2}|F^{s_1}_{\lambda_3,\lambda_4}|^2d\Omega_1d\bar\Omega_2,
\end{eqnarray}
where we replace the $\lambda_1$ helicity projection with $\lambda'_1$ for resonance $s_1$ in the $\mathcal{A}_\textrm{I}^{*}(...)$ amplitude, and its Breit-Wigner function is not included for simplicity.

The above equation indicates that the decay rate calculated in the canonical states is equal to that calculated in the helicity states, if the helicity amplitude is taken as

\begin{eqnarray}
&&\mathcal{H}_\textrm{I}(\theta_1,\bar\theta_2,\phi_1,\bar\phi_2;M,\lambda_2,\lambda_3,\lambda_4)\nonumber\\
&=&a\sum_{\lambda_1}\djmn{J*}{M,\lambda_1-\lambda_2}{\phi,\theta_1,0}\djmn{s_1*}{\lambda_1,\lambda_3-\lambda_4}{\bar\phi_2,\bar\theta_2,0}\nonumber\\
&\times&F^{J}_{\lambda_1,\lambda_2}F^{s_1}_{\lambda_3,\lambda_4},
\end{eqnarray}
where $a$ is a normalization factor.

\subsection{Coherent amplitudes for double decay chains}
\begin{table*}
\caption{Variables defined for the decay chain II, $J\to s_0+s_3$, with $s_0\to s_2+s_4$. \label{chainb}}
\begin{tabular}{lccc|lccc}
\hline\hline
 &\multicolumn{3}{c}{First decay} & \multicolumn{3}{|c}{Second decay} \\\hline
      & Parent & Daughter 1 & Daughter 2& Parent & Daughter 1 & Daughter 2\\\hline
Spin  &  $J$   & $s_0$      & $s_3$&$s_0$&$s_2$&$s_4$\\
Spin $z$ projection & $M$ & $m_0$ & $m_3$&$m_1$&$m_2$&$m_4$\\
Helicity &---             & $\lambda_0$ & $\lambda^3$&$\lambda_0\lambda'_0$&$\lambda^2$&$\lambda^4$\\
Helicity amplitude & &$F^J_{\lambda_0,\lambda^3}$ &   & &$F^{s_0}_{\lambda^2,\lambda^4}$&&\\
Solid angle &  & $\Omega^1(\theta^1,\phi^1)$ & & & $\Omega^2(\theta^2,\phi^2)$& \\
\hline\hline
\end{tabular}
\end{table*}

The amplitude in the canonical states for the decay chain II is
\begin{eqnarray}
&&\mathcal{A}_\textrm{II}(\theta^1,\theta^2,\phi^1,\phi^2; J,M,m_2,m_3,m_4)\nonumber\\
&=&\sum_{\lambda_0,\lambda^2,\lambda^3,\lambda^4}
\djmn{s_2}{m_2,\lambda^2}{\phi^2,\theta^2,0}
\djmn{s_3}{m_3,-\lambda^3}{\phi^1,\theta^1,0}\nonumber\\
&\times&\djmn{s_4}{m_4,-\lambda^4}{\phi^2,\theta^2,0}\djmn{J*}{M,\lambda_0-\lambda^3}{\phi^1,\theta^1,0}\nonumber\\
&\times&\djmn{s_0*}{\lambda_0,\lambda^2-\lambda^4}{\bar\phi^2,\bar\theta^2,0}
F^J_{\lambda_0,\lambda^3}F^{s_0}_{\lambda^2,\lambda^4},
\end{eqnarray}
where $\bar\Omega^2(\bar\theta^2,\bar\phi^2)$ are the helicity angles, by which we rotate the $z$ axis of $s_0$ helicity frame into the direction of $s_2$ momentum defined in their mother particle rest frames. The total amplitude is obtained by adding the two chains coherently. To express it in the canonical basis with solid angles ($\theta_1,\theta_2,\phi_1,\phi_2$), we split the spin rotations in chain II into two successive rotations. First, we rotate the $z$ axis of spin $s_i$ into the direction of momentum defined in chain I, with solid angles $\theta_1/\theta_2$ , and then rotate it into the direction of the momenta defined in chain II, by $\theta^1_2/\theta^2_{1,2}$ with these relations
\begin{eqnarray}\label{angrelation}
\djmn{s_3}{m_3,-\lambda^3}{\phi^1,\theta^1,0}=\sum_{k_3}\djmn{s_3}{m_3,k_3}{\phi_2,\theta_2,0}\djmn{s_3}{k_3,-\lambda^3}{0,\theta^1_2,0},\nonumber\\
\djmn{s_2}{m_2,\lambda^2}{\phi^2,\theta^2,0}=\sum_{k_2}\djmn{s_2}{m_2,k_2}{\phi_1,\theta_1,0}\djmn{s_2}{k_2,\lambda^2}{0,\theta^2_1,0},\nonumber\\
\djmn{s_4}{m_4,-\lambda^4}{\phi^2,\theta^2,0}=\sum_{k_4}\djmn{s_4}{m_4,k_4}{\phi_2,\theta_2,0}\djmn{s_4}{k_4,-\lambda^4}{0,\theta^2_2,0}\nonumber.\\
\end{eqnarray}
In the above equations, no azimuthal rotations are needed to align the $s_i$ helicities in two decay chains, since the decay planes are the same.

The decay rate corresponding to the total amplitude reads
\begin{eqnarray}
d\Gamma&\propto& d\Omega_1d\bar\Omega_2\sum_{\bar M,m_2,m_3,m_4}|\mathcal{A}_\textrm{I}(...)|^2+|\mathcal{A}_\textrm{II}(...)|^2\nonumber\\
&+&\mathcal{A}_\textrm{I}(...)\mathcal{A}^*_\textrm{II}(...)+
\mathcal{A}^*_\textrm{I}(...)\mathcal{A}_\textrm{II}(...),
\end{eqnarray}
where $(...)$ denotes $(\theta_1,\theta_2,\phi_1,\phi_2;M,m_2,m_3,m_4)$, and Eq. (\ref{angrelation})
has been used to make a replacement in the amplitude $\mathcal{A}_\textrm{II}$.
Calculation of interference terms is straightforward. Using the orthogonal relations of $D$-functions and summing over the spin projections $m_2,m_3$, and $m_4$, one has

\begin{eqnarray}
&&\sum_{\bar M,m_2,m_3,m_4}\mathcal{A}_\textrm{I}(...)\mathcal{A}^*_\textrm{II}(...)\nonumber\\
&=&\sum_{\lambda_1,\lambda_2,\lambda_3,\lambda_4}\djmn{J*}{M,\lambda_1-\lambda_2}{\phi,\theta_1,0}
\djmn{s_1*}{\lambda_1,\lambda_3-\lambda_4}{\bar\phi_2,\bar\theta_2,0}\nonumber\\
&\times&F^{J}_{\lambda_1,\lambda_2}F^{s_1}_{\lambda_3,\lambda_4}\left[\sum_{\lambda_0,\lambda^2,\lambda^3,\lambda^4}
\djmn{J}{M,\lambda_0-\lambda^3}{\phi^1,\theta^1,0}\right.\nonumber\\
&\times&\djmn{s_0}{\lambda_0,\lambda^2-\lambda^4}{\bar\phi^2,\bar\theta^2,0}F^{J}_{\lambda_0,\lambda^3}F^{s_0}_{\lambda^2,\lambda^4}d^{s_2}_{\lambda^2,\lambda_2}(\theta^2_1)\nonumber\\
&\times&d^{s_3}_{\lambda^3,-\lambda_3}(\theta^1_2)d^{s_4}_{\lambda^4,-\lambda_4}(\theta^2_2)\left]\right.,
\end{eqnarray}
where the rotations, $d^{s_2}_{\lambda^2,\lambda_2}(\theta^2_1),~d^{s_3}_{\lambda^3,-\lambda_3}(\theta^1_2)$ and $d^{s_4}_{\lambda^4,-\lambda_4}(\theta^2_2)$, transform the helicities $\lambda^2,~\lambda^3$ and $\lambda^4$ defined in chain II, into those defined in chain I, respectively. However, these rotations are canceled in the calculation of the term $\sum_{\bar M,m_2,m_3,m_4}|\mathcal{A}_\textrm{II}(...)|^2$ due to the orthogonal relations of the Wigner-$d$ function. Hence, the coherent amplitude in helicity bases for the two chains is taken as

\begin{eqnarray}\label{mastereq}
&&\mathcal{H}(M,\lambda_2,\lambda_3,\lambda_4)\nonumber\\
&=&a~BW(M_1)\sum_{\lambda_1}\djmn{J*}{M,\lambda_1-\lambda_2}{\phi_1,\theta_1,0}\nonumber\\
&\times&\djmn{s_1*}{\lambda_1,\lambda_3-\lambda_4}{\bar\phi_2,\bar\theta_2,0}F^{J}_{\lambda_1,\lambda_2}F^{s_1}_{\lambda_3,\lambda_4}\nonumber\\
&+&b~BW(M_0)\sum_{\lambda_0,\lambda^2,\lambda^3,\lambda^4}
\djmn{J*}{M,\lambda_0-\lambda^3}{\phi^1,\theta^1,0}\nonumber\\
&\times&\djmn{s_0*}{\lambda_0,\lambda^2-\lambda^4}{\bar\phi^2,\bar\theta^2,0}F^{J}_{\lambda_0,\lambda^2}F^{s_0}_{\lambda^2,\lambda^4}\nonumber\\
&\times&d^{s_2}_{\lambda^2,\lambda_2}(\theta^2_1)d^{s_3}_{\lambda^3,-\lambda_3}(\theta^1_2)d^{s_4}_{\lambda^4,-\lambda_4}(\theta^2_2),
\end{eqnarray}
where $a$ and $b$ are coupling constants, $BW(M_0)$ and $BW(M_1)$ are the Breit-Wigner functions for resonances $s_0$ and $s_1$, respectively. To coherently add the third decay chain to the amplitude, e.g., $J\to s_4+s_5$ with $s_5\to s_2+s_3$, the generation of the above formula is straightforward by multiplying its helicity sequential decay amplitude with the rotations to transform the helicities defined in this chain to those defined in chain I.
\section{Illustrative examples\label{sectionIII}}
Here we give two examples to illustrate the principle to construct the helicity amplitudes with coherent interference effects. We confine ourselves to the case of three-body decays with two pseudoscalar mesons in the final states.

\subsection{$e^+e^-\to\gamma^*\to \pi^+\pi^-\jp$}
The charmoniumlike state, $Z_c(3900)^\pm$, was observed for the first time by the BESIII Collaboration in the process $e^+e^-\to \pi^+\pi^-\jp$, and confirmed by the Belle and CLEO Collaborations \cite{bes3zc,bellezc,cleoczc}. We consider two kinds of decays in this process,
\begin{eqnarray}
\textrm{I}&:&\ee\to \gamma^* \to f_0(980) \jp(\lambda_1)~(\phi_1,\theta_1)\nonumber\\
&&\textrm{~with~} f_0(980)\to\pi^+\pi^-,\nonumber\\
\textrm{II}&:&\ee\to\gamma^*\to \pi^\pm Z_c(3900)^\mp(\lambda_2)~(\phi_2,\theta_2)\nonumber\\
&&\textrm{~with~} Z_c(3900)^\pm\to\pi^\pm\jp(\lambda_3)~(\phi_3,\theta_3).\nonumber
\end{eqnarray}
Here $\lambda_i(i=1,2,3)$ are the helicity values for the corresponding particles, and $\theta_i$ and $\phi_i$ are the polar and azimuthal angles defined in the helicity reference for each decay, respectively. We assume the spin and parity of $Z_c(3900)$ to be $1^+$. The coherent helicity amplitude reads
\begin{eqnarray}
\mathcal{H}&=&a~BW(f_0,m_{\pi^+\pi^-})\djmn{1*}{M,\lambda_1}{\phi_1,\theta_1,0}F^{\gamma*\to \jp f_0}_{\lambda_1,0}\nonumber\\
&+&b~BW(Z_c^\pm,m_{\pi^\mp\jp})\sum_{\lambda_2,\lambda_3}\djmn{1*}{M,\lambda_2}{\phi_2,\theta_2}\nonumber\\
&\times&\djmn{1*}{\lambda_2,\lambda_3}{\phi_3,\theta_3,0}d^{1}_{\lambda_3,\lambda_1}(\theta^3_1)F^{\gamma^*\to Z_c\pi}_{\lambda_2,0}F^{Z_c\to\jp \pi}_{\lambda_3,0},\nonumber\\
\end{eqnarray}
where $M=\pm1$ is the $z$ projection of $\gamma^*$ spin, $a$ and $b$ are coupling constants, $\theta^3_1$ is the angle between the momenta of $\jp$ in the $\ee$ CM system and $Z_c^\pm$ rest frame, $BW$ denotes Breit-Wigner function, $F$s are the helicity amplitudes and are reduced by relating them to the $LS$ coupling amplitudes as \cite{chung1,chung2,chung3}
\begin{eqnarray}
F^{\gamma^*\to \jp f_0}_{-1,0}&=&F^{\gamma^*\to \jp f_0}_{1,0}={g^a_{0,1}\over \sqrt 3}+{g^a_{2,1}r_a^2\over \sqrt 6},\nonumber\\
F^{\gamma^*\to \jp f_0}_{0\,0}&=&{\gamma^a_sg^a_{0,1}\over \sqrt 3}-\sqrt{2\over 3}{\gamma^a_sg^a_{2,1}r_a^2\over \sqrt 6}\nonumber\\
F^{\gamma^*\to Z_c\pi}_{-1,0}&=&F^{\gamma^*\to Z_c\pi}_{1,0}={g^b_{0,1}\over \sqrt 3}+{g^b_{2,1}r_b^2\over \sqrt 6},\nonumber\\
F^{\gamma^*\to \jp f_0}_{0\,0}&=&{\gamma^b_sg^b_{0,1}\over \sqrt 3}-\sqrt{2\over 3}{\gamma^b_sg^b_{2,1}r_b^2\over \sqrt 6}\nonumber\\
F^{ Z_c\to\jp\pi}_{-1,0}&=&F^{Z_c\to\jp\pi}_{1,0}={g^c_{0,1}\over \sqrt 3}+{g^c_{2,1}r^2\over \sqrt 6},\nonumber\\
F^{\gamma^*\to \jp f_0}_{0\,0}&=&{\gamma^c_sg^c_{0,1}\over \sqrt 3}-\sqrt{2\over 3}{\gamma^c_sg^c_{2,1}r_c^2\over \sqrt 6}.
\end{eqnarray}
where $g^i_{l,S}(i=a,b,c)$ are coupling constants, $r_i(i=a,b,c)$ is the magnitude of breakup momentum of the two-body decays. $\gamma^i_s(i=a,b,c)$ is the ratio of $\jp$ energy to its mass in the decay.
In these decays, parity conserves the helicity amplitudes.

\subsection{$\ldc\to pK^-\pi^+$}
The $\ldc$ has a sizable branching fraction ($5.0\pm1.3$)\% decaying into the $pK^-\pi^+$ final states. Amplitude analysis is desirable to extract the resonance contributions to this decay, such as $\bar K^*(892)^0,~\Delta(1232)^{++}$, excited $\Lambda$ and $\Sigma$ states.  We consider three types of decay like
\begin{eqnarray}
&\textrm{I}:& \ldc\to p(\lambda_1)\bar K^*(892)^0(\lambda_2) ~(\phi_1,\theta_1),\nonumber\\
&&\bar K^*(892)^0\to K^-\pi^+~(\phi_2,\theta_2);\nonumber\\
&\textrm{II}:& \ldc\to \Delta(1232)^{++}(\lambda_3)K^-~(\phi_3,\theta_3),\nonumber\\
&&\Delta(1232)^{++}\to p(\lambda_4)\pi^+~(\phi_4,\theta_4);\nonumber\\
&\textrm{III}:&\ldc\to\Lambda(1520)(\lambda_5)\pi^+~(\phi_5,\theta_5),\nonumber\\
&&\Lambda(1520)\to p(\lambda_6) K^-~(\phi_6,\theta_6),
\end{eqnarray}
where $\lambda_i(i=1,...,6)$ are helicity values for corresponding particles, and $\theta_i$ and $\phi_i$ are the polar and azimuthal angles defined in the helicity reference system for each decay.
For the decay I, the helicity amplitude reads
\begin{eqnarray}
\mathcal{H}_I&\propto& BW(K^*,m_{K^-\pi^+})\sum_{\lambda_2} \djmn{\half*}{M,\lambda_1-\lambda_2}{\phi_1,\theta_1,0}\nonumber\\
&\times&\djmn{1*}{\lambda_2,0}{\phi_2,\theta_2,0}F^{\ldc\to pK^*}_{\lambda_1,\lambda_2}F^{K^*\to K\pi}_{0,0},
\end{eqnarray}
where $M$ is the $z$ projection of $\ldc$ and $F$s are the helicity amplitudes.

For the decay II, the helicity amplitude reads
\begin{eqnarray}
\mathcal{H}_\textrm{II}&\propto& BW(\Delta^{++},m_{p\pi^+})\sum_{\lambda_3,\lambda'_4}\djmn{\half*}{M,\lambda_3}{\phi_3,\theta_3}\nonumber\\
&\times&\djmn{{3\over 2}*}{\lambda_3,\lambda'_4}{\phi_4,\theta_4}F^{\ldc\to\Delta K}_{\lambda_3,0}F^{\Delta\to p\pi}_{\lambda'_4,0}d^{\half}_{\lambda'_4,\lambda_1}(\theta^4_1),\nonumber\\
\end{eqnarray}
where $\theta^4_1$ is the angle between the momenta of proton in $\Delta^{++}$ and $\ldc$ rest frames.

For the decay III, the helicity amplitude reads
\begin{eqnarray}
\mathcal{H}_\textrm{III}&\propto& BW(\Lambda,m_{pK^-})\sum_{\lambda_5,\lambda'_6}\djmn{\half*}{M,\lambda_5}{\phi_5,\theta_5,0}\nonumber\\
&\times&\djmn{\half*}{\lambda_5,\lambda'_6}{\phi_6,\theta_6,0}d^{\half}_{\lambda'_6,\lambda_1}(\theta^6_1)\nonumber\\
&\times&F^{\ldc\to\Lambda\pi^+}_{\lambda_5,0}F^{\Lambda\to pK^-}_{\lambda'_6,0},
\end{eqnarray}
where $\theta^6_1$ is the angle between the momenta of proton calculated in $\Lambda$ and $\ldc$ rest frames.

The coherent helicity amplitude is taken as
\begin{equation}
\mathcal{H}=c_1\mathcal{H}_\textrm{I}+c_2\mathcal{H}_\textrm{II}+c_3\mathcal{H}_\textrm{III},
\end{equation}
where $c_1,~c_2$, and $c_3$ are coupling constants.

The helicity amplitudes of $F$'s functions are reduced by relating them to the $LS$ coupling amplitudes as
\begin{eqnarray}
F^{\ldc\to pK^*}_{\half,1}&=&-F^{\ldc\to pK^*}_{-\half,-1}\nonumber\\
&=&{Wg_{0,\half}\over \sqrt 3}-{g_{1,{3\over 2}}r\over \sqrt 6}+{g_{1,\half}r\over \sqrt 3}-{Wg_{2,{3\over 2}}r^2\over \sqrt 6},\nonumber\\
F_{0,0}^{K^*\to K\pi}&=&rg_{0,1},\nonumber
\end{eqnarray}
\begin{eqnarray}
F^{\ldc\to\Delta K}_{\half,0}&=&{r^2W\gamma_{\Delta}g_{2,{3\over 2}}\over \sqrt 2}-{r\gamma_{\Delta}g_{1,{3\over 2}}\over \sqrt 2},\nonumber\\
F^{\ldc\to\Delta K}_{-\half,0}&=&-{1.5r^2W(\gamma_{\Delta}^2+1)g_{2,{3\over 2}}\over 3\sqrt 2}-{1.5r(\gamma^2_{\Delta}+1)g_{1,{3\over 2}}\over3 \sqrt 2},\nonumber\\
F^{\Delta\to p\pi}_{\half,0}&=&-F^{\Delta\to p\pi}_{-\half,0}=-{r^2Wg_{2,\half}\over \sqrt 2},\nonumber\\
F^{\ldc\to\Lambda\pi^+}_{\half,0}&=&{g_{0,\half}\over \sqrt 2}-{rWg_{1,\half}\over \sqrt 2},\nonumber\\
F^{\ldc\to\Lambda\pi^+}_{-\half,0}&=&{\gamma_{\Lambda}g_{0,\half}\over \sqrt 2}+{rW\gamma_{\Lambda}g_{1,\half}\over \sqrt 2},\nonumber\\
F^{\Lambda\to pK^-}_{\half,0}&=&{g_{0,\half}\over \sqrt 2},\nonumber\\
F^{\Lambda\to pK^-}_{\half,0}&=&{\gamma_pg_{0,\half}\over \sqrt 2},
\end{eqnarray}
where $g_{LS}$ is the coupling constant, $\gamma_x$ is the ratio of energy to mass of the $x$ particle in the decay, and $r$ is the magnitude of breakup momentum for the two-body decay. In the $\ldc$ decays, the parity doesn't conserve; therefor all possible waves of orbital momentum are included. For $K^*,~\Delta$, and $\Lambda(1520)$ decays, the parity conserves the amplitudes.
\section{Conclusion and discussion}
The method of helicity amplitude is widely used in the partial wave analysis. From the viewpoint of the experiment side, a few resonances are introduced to model the production of final states. The helicities of final states are defined along the direction of outgoing particle in the rest frame of their mother particle system. If there are different decay chains involved, this makes the sum of the amplitudes inconsistent since
the helicity of the same particle may have different definitions. One has to introduce an additional rotation to transform the helicity into the same reference. We present a deviation of coherent helicity amplitude for the three-body decays. The principle can be generalized to other cases, e.g., four-body decays.

If the amplitudes for multichain decays are constructed in the canonical basis, the spins of particles involved are defined in the same reference. This allows one to add them coherently. We borrow this idea in our derivation by relating the amplitude defined in the helicity base to that defined in the canonical basis. In the examples of three-body decays, we show that the helicity amplitudes need additional rotations to allow them to add coherently [see Eq. (\ref{mastereq})]. These rotations may have significant impact on the interference between the different decay chains.

Two examples are shown to construct the coherent helicity amplitudes for the case of three-body decays. For practical purposes, the formulas are further reduced by using the covariant helicity-coupling amplitudes. It is important to note that these additional rotations are unneeded if the intermediate states are introduced with the same decay sequence topology. Since the intermediate states are reconstructed with the same final states, their helicities are defined in the same reference.

\vspace{1cm} {\bf Acknowledgements:} The author acknowledges suggestions by Professor C. P. Shen. The work is partly supported by
the National Natural Science Foundation of China under Grants No. 11645002,
11375205, and 11565006.

\end{document}